# A Comment on the note arXiv: 2006.13147 on arXiv:2005.05301, 'Preparation of the Neutrino-4 experiment on search for sterile neutrino and the obtained results of measurements'

A.P. Serebrov*, R.M. Samoilov (on behalf of Neutrino-4 collaboration)

*NRC "KI" Petersburg Nuclear Physics Institute, Gatchina, Russia*

Abstract

Here is response Neutrino-4 collaboration to the note arXiv:2006.13147 on article "Preparation of the Neutrino-4 experiment on search for sterile neutrino and the obtained results of measurements" arXiv:2005.05301. Red text is commentary from the note arXiv:2006.13147, black text is our answers.

The reply to your comment in Section 2 on using a method $\Delta\chi^2$ contains in Section 22 of our article.

"It is often discussed that stricter limitations on the confidence level of the result can be obtained using the Feldman-Cousins method. In compliance Wilks theorem $\Delta\chi^2$ method is possible to apply successfully if effect is observed at the level of reliability 3σ more. The result of processing without taking into account systematic errors with an energy interval of 500 keV is $\sin^2 2\theta_{14} = 0.38 \pm 0.11(3.5\sigma)$, and when averaging data over 125keV, 250keV and 500keV is $\sin^2 2\theta_{14} \approx 0.26 \pm 0.08(3.2\sigma)$. Since the reliability of the effect we observe exceeds $3\sigma$, we do not consider it mandatory to use the Feldman-Cousins method and propose to do another additional analysis of our data.

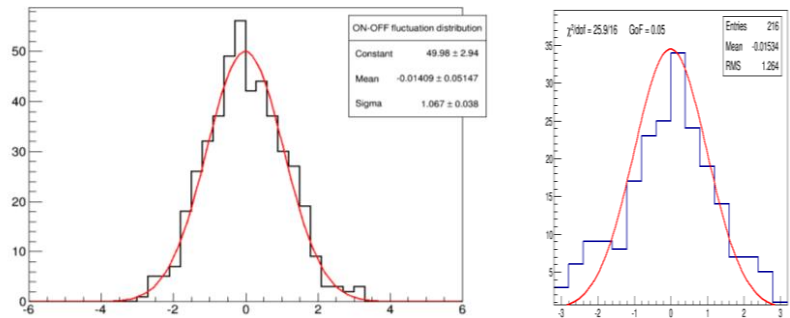

FIG.49. Top - distribution of the count rate ON- OFF in the entire energy range, normalized by $\sigma$. Bottom - distribution $R_{ik}^{\text{exp}}$ of all 216 points over the L/E range from 0.9 to 4.7, normalized by $\sigma$.

Initial distribution of the count rate ON- OFF in the entire energy range was shown in Figure 34 (bottom) and in Fig. 49 (top) (*hereinafter, the numbering of the figures is taken from the original article*) for obviousness. It shows a normal distribution determined practically by statistics. In fig. 49, we compare it with the distribution obtained for the ratio $R_{ik}^{\text{exp}}$ for the same dataset. It, as well as the distribution ON-OFF, normalized by $\sigma$. Figure 49 (bottom) shows the distribution of all 216 points over the L/E range from 0.9 to 4.7. You can see that the distribution $R_{ik}^{\text{exp}}$ already differs from normal ($\sigma = 1$, $\mu = 0$ and it normalized as $R_{ik}^{\text{exp}}$) due to the effect of oscillations. Value of the $\chi^2/\text{dof}$ parameter is 25.9/16 which disfavors this function because of confidence level for this result is only 5%. It can be seen that this analysis uses initial data before process of summing up to obtain oscillation parameters. We would like to note that the effect of oscillations is manifested using three processing methods.

1. $\Delta\chi^2$ method at plane $(\sin^2 2\theta_{14}, \Delta m_{14}^2)$ for $R_{ik}^{\text{exp}}$ (but not for $\text{ON} - \text{OFF} = N(E_i, L_k)$ to avoid spectral dependence.

2. Coherent summation method by variable L/E,

3. Analysis of distribution $R_{ik}^{\text{exp}}$ as opposed to normal distribution due to the effect of oscillations."

---

* serebrov_ap@pnpi.nrcki.ru

Our article, which the commentators criticize, says on page 20:
«The stability of the results of the analysis can be tested. Using the obtained experimental data ($N_i, \pm \Delta N_i$,) one can perform a data simulation using randomization with a normal distribution around $N_{i,k}$ with dispersion $\Delta N_{i,k}$. Applying this method, 60 virtual experiments were simulated with results lying within current experimental accuracy. One can carry out the analysis described above for virtual experiments and average results over all distributions. It was observed that exclusion area (pink area in Fig. 10a) coincide with experimental one and oscillation effect area is gathered around value $\Delta m_{14}^2 \approx 7.3 eV^2$. Finally, one can simulate the experimental results with same accuracy but in assumption of zero antineutrino oscillations. Obtained result reveals that amplitude of perturbations in horizontal axes, i.e. values of $\sin^2 2\theta_{14}$, is significantly reduced. It signifies that big perturbations in Fig. 10a indicate an existence of the oscillation effect. Simulated experimental data distributions with same accuracy, but in assumption of zero oscillation allows us to estimate sensitivity of the experiment at CL 95% and 99%. Obtained estimations can be used to compare our results with other experiments.»
Please, read more carefully the article that you criticize.

For a better understanding of the above, figures 55 is added with two pictures and present the results of this analysis with and without the assumption of the existence of the oscillation effect. Recall that the simulations were performed with the same statistical accuracy as in the experiment, but repeatedly (60 times). Figure 55 (on left) shows that if there is an effect with the parameters found in our experiment, we get confirmation of the effect in the same area. In addition, satellites appear as additional evidence of the presence of an effect.

In section 18 on page 20 we have this statement:
"The satellites appear due to effect of harmonic analysis where in presence of noises along with base 6 frequency we also can obtain frequencies equal to base frequency multiplied by integers and half-integers."

Figure 55 (in the middle) shows that «in assumption of zero oscillation allows us to estimate sensitivity of the experiment at CL 95% and 99%. Obtained estimations can be used to compare our results with other experiments.» as was stated in the same section. This comparison with much larger statistics in the next stage of the experiment was presented in figure 55 (on the right). It shows that the sensitivity of the Neutrino-4 experiment is twice the sensitivity of the STEREO and POSPECT experiments.

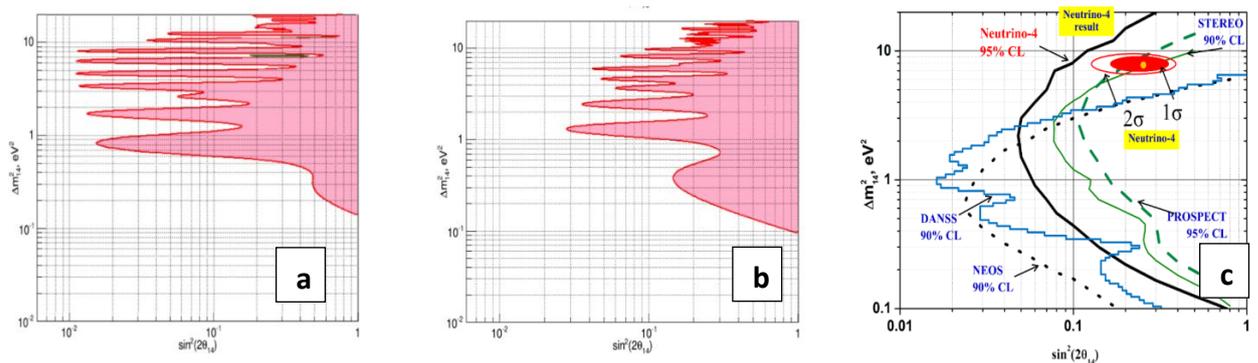

Figure 55. a,b – MC-based statistical approach in the Neutrino-4 published results; c – Comparison of planes of parameters (E,L) in experiments Neutrino-4, STEREO and PROSPECT.

**Finally, note that section 15 presents a complete MC simulation of the experiment with very high statistical accuracy to demonstrate what should be expected in such data processing. The resulting picture unambiguously indicates the expediency of using the coherent summation method by variable L/E.**

"Model of the experimental setup and MC simulation of experimental data together with suggested method of data analysis applied to simulated data reveal how the oscillation effect should manifest itself at E,L plane. In this section we present results of MC simulation in which we incorporated geometric configuration of the antineutrino source and detector including the sectioning. In this simulation we have already used optimal parameters $\Delta m_{14}^2$ and $\sin^2 2\theta_{14}$, which were derived from the analysis of experimental data.

The source of antineutrino with geometrical dimensions of the reactor core 42x42x35cm³ was simulated, as well as a detector of antineutrino taking into account its geometrical dimensions (50 sections of 22.5x22.5x75cm³). The antineutrino spectrum of U²³⁵ (though it does not matter since energy spectrum in equation (2) is cancelled out) factored by function of oscillations $1 - \sin^2 2\theta_{14} \sin^2(1.27\Delta m_{14}^2 L_k/E_i)$ was used.

The most important parameter in this simulation was the energy resolution of the detector (2σ), which was set to be ±250 keV. Fig. 33 (left) illustrates the simulated matrix of ratio $N_{ik}L_k^2/K^{-1}\sum N_{ik}L_k^2$ which we suggest to use for data analysis based on equation (2). In simulation the statistical accuracy of ratio $\Delta N_{ik}/N_{ik}$ equal to 1%, which is significantly better than the experimental value. The MC simulation can be summarized in several conclusions. First of all, it becomes obvious what pattern the expected oscillation effect can bring at E, L plane. It also reveals that data from equal L/E should be summarized to carry out analysis. Indeed, such analysis can result in oscillation curve which is shown in fig.33 on the right. The next important conclusion is that oscillation is fading and that effect depends on energy resolution of the detector. One can compare to examples with energy resolutions of the detector being ± 125 keV и ±250 keV which is shown in fig. 33. Note that the energy resolution of the detector determines the number of observed oscillations, but not the amplitude.

Therefore, the data analysis should be carried out by calculating ratio $N_{ik}L_k^2/K^{-1}\sum N_{ik}L_k^2$ for different points at E, L plane and then summing up the values for points with equal ratio L/E. It should be noted that integration of the matrix over energy or distance significantly suppresses the ability to detect the effect of oscillations.

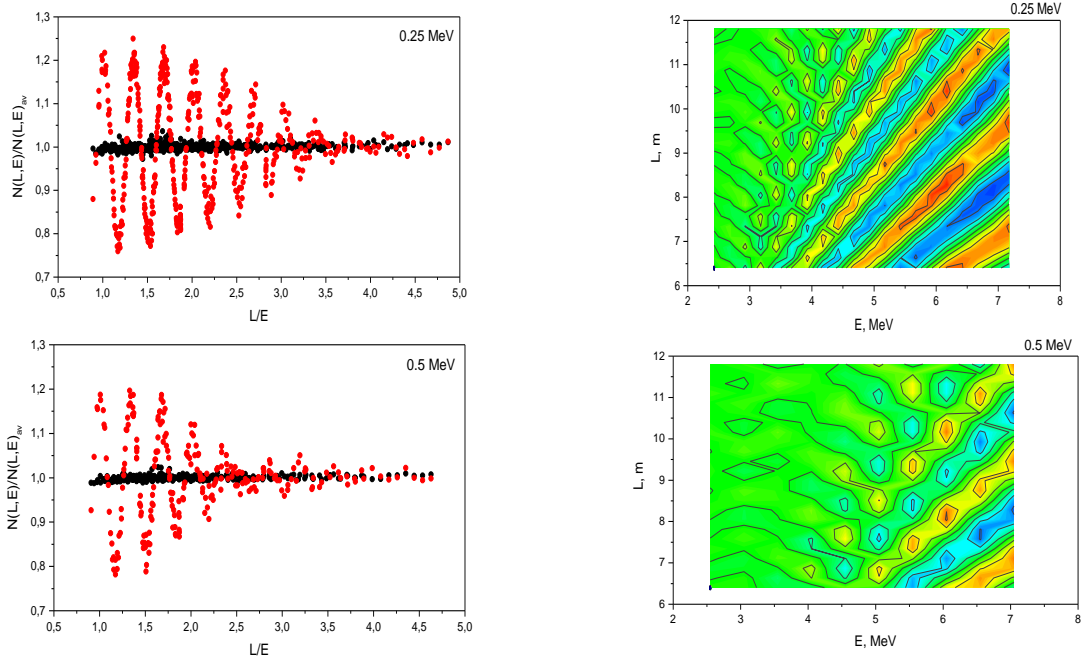

FIG.33. Modeling of the ratio $N_{ik}L_k^2/K^{-1}\sum N_{ik}L_k^2$ at E,L plane with different energy resolutions of the detector $2\sigma = 250$ keV and $2\sigma = 500$ keV (on the left side), ratio $N_{ik}L_k^2/K^{-1}\sum N_{ik}L_k^2$ as dependence from E/L (on the right side). Red dots are for oscillation, black dots are for absence of oscillation.

As you can see answers to your remarks is already in our article.

Reply to section 3, which states: «In the previous section, we demonstrated that statistical fluctuations could be mistaken for an oscillation signature in absence of a proper statistical approach. The problem could be exacerbated if there are unidentified, and thus unaccounted for, oscillation-mimicking systematic effects. This can be especially true for an experiment with a small detector located in close proximity to a reactor with little overburden. Here we point out two such key systematic effects which are not discussed by the Neutrino-4 experiment.

Neutrino-4 uses gadolinium-doped liquid scintillator (LS) as the target and inverse beta decay (IBD) mechanism for detecting neutrinos. The energy of the positron produced in the IBD interaction acts as a proxy for the $\nu_e$ energy while the neutron produced in the interaction captures on Gadolinium producing 3 to 4 Mev gamma rays and is used to establish a coincident signal. In a segmented detector of scale ~1 meter like Neutrino-4, the IBD positron and/or the annihilated s lose some of their energy through escape or energy deposition in inactive volume. This leads to an energy spectrum that is position-dependent since the IBD interactions taking place close to the edge of the detector have a higher fraction of escape energy. Additionally, since the attenuation length of high energy s produced from neutron capture on Gadolinium is 15 cm, the neutron capture efficiency also varies within the detector volume. These edge effects could induce complex correlations between energy and efficiency which could induce an oscillation-like signature. The situation is further complicated by the fact that the detector is mobile and each detector segment spans over multiple baselines. However, these complex detector effects can be accounted for in the oscillation search by using a fully validated detector MC simulation. While it is not clear that these concerns introduce false oscillations, they will certainly effect the estimation of signal significance, and must be accounted for in a full analysis. There is no indication that Neutrino-4 has incorporated their detector MC simulation in the theoretical rates from the Eq. 1. Therefore, we request Neutrino-4 experiment to provide more details on the calculation of their theoretical rates alongside a detector MC simulation that is fully validated using calibration data.»

First of all, we consider it necessary to indicate that the MC model of the detector is presented in sufficient detail in section 11 on pages 14-15. All the detection features mentioned in the commentary were considered for the purpose of calculating the effectiveness of the detector as a whole, taking into account its sectional structure. The result was shown in Figure 27. This was done in order to obtain information about the sensitivity of the experiment.

**However, it should be explained that we use a different approach, understanding the extremely high complexity of considering all these features to determine the absolute efficiency of the detector.**

Here, as in a number of other experiments, we use the method of relative measurements. The problems mentioned herein are solved by using relative measurements with the same detector at different distances as well as measurements by means different sections of the detector at the same distance. Scheme of our installation has been shown in Figure 18. This is the reason that the scheme shown below was selected. The stationary detector approach is not optimal. At the present time STEREO and PROSPECT experiments are using stationary detector scheme and the questions mentioned in your note have more relations to them.

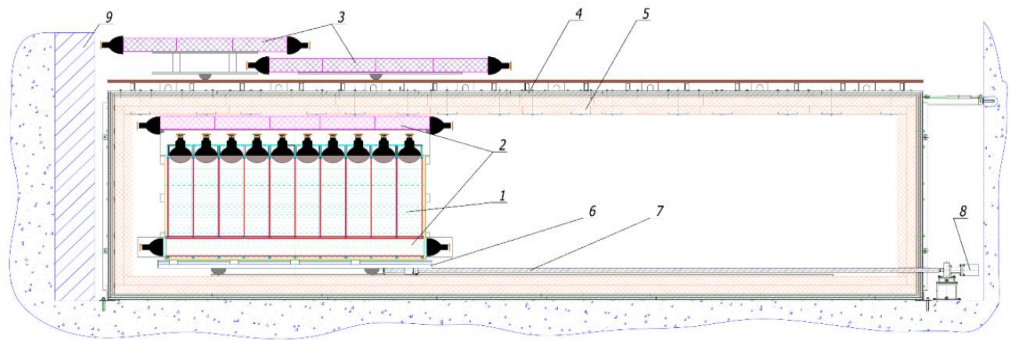

FIG. 18. General scheme of an experimental setup. 1 – detector of reactor antineutrino, 2 – internal active shielding, 3 – external active shielding (umbrella), 4 – steel and lead passive shielding, 5 – borated polyethylene passive shielding, 6 – moveable platform, 7 – feed screw, 8 – step motor, 9 – shielding against fast neutrons made of iron shot.

The absolute efficiency of its individual sections is not a problem precisely due to the measurement method with the movement of the detector. The different efficiency of different sections in one row (edge and center) in connection with the escape of gadolinium or annihilation gamma-quanta is nevertheless the same for different rows except the first and last. They are not used for recording positron signals, but are used as gamma catcher, active and passive shielding.

The idea of moving the detector is extremely important, since different rows appear at the same distances. This averages the difference in calibration across 40 sections and 8 rows. This is the most important advantage of the Neutrino-4 detector in relation to stationary detectors. All these issues were discussed in detail in section 20 on pages 22-23 and in figure 43: «To consider how differences in rows efficiencies affect the final results, one must take into account that averaging of spectra obtained with various rows at the same distance. In that approach the square deviation from the mean value is ~ 2.5%, as shown in figure 43. It indicates that the influence of detector inhomogeneity on the L/E dependence is insignificant and cannot be the origin of oscillation effect.»

Please, read more carefully the article that you criticize.

Also, section 3 says:

> «Short baseline reactor neutrino experiments have to overcome challenging—often position- and energy-dependent— background environments in the search for $\bar{\nu}_e$ oscillations. The correlated cosmogenic backgrounds can be measured during the reactor-off period and can be scaled to reactor-on period and subtracted from the reactor-on data. In principle, this works well if the detector has a good signal-to background (S:B) ratio, the reactor-off duration accounts for a significant amount (~50 %) of data-taking, and there exists no variations 2 in cosmogenic backgrounds between reactor on and -off periods. As discussed on page 8 in Ref. [10], in an earlier analysis, a fit performed using only 278 days of reactor-off data was found to yield an oscillation parameter at ~$2.6\sigma$ C.L. This could be an artifact of assigning incorrect significance based on using standard $\chi^2$ as discussed in the previous section. Conversely, in conjunction with the low S:B ratio of ~ 0:5 of Neutrino-4 experiment [1], it could also suggest the possibility that the oscillation signature indicated by fluctuations in the cosmogenic backgrounds gets enhanced with addition of reactor-on data. It is also possible that this is unrelated to the oscillation signature suggested by the IBD data. To disambiguate between various scenarios, we encourage the Neutrino-4 collaboration to provide more details on backgrounds and the background subtraction procedure employed in calculating the experimental IBD spectrum shown in Eq. 2.»

First of all, when discussing background problems, it should be notes that the oscillations of the cosmic background have been studied in detail and was presented in Figure 8 on page 7 and for convenience are shown below

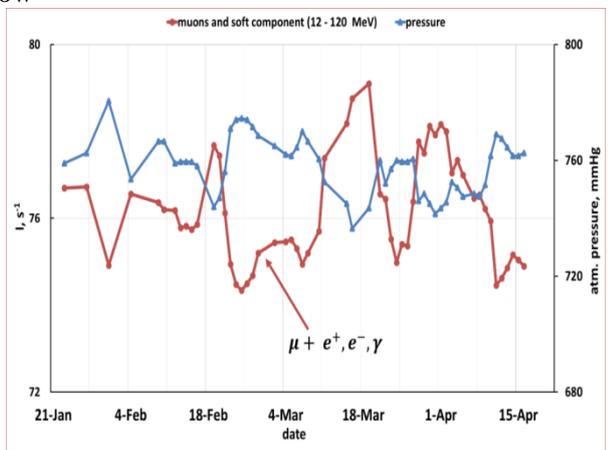

FIG. 8. Barometric effect of cosmic rays: the left axis illustrates a summary detector count rate in the energy areas 3 and 4, the right axis shows atmospheric pressure, the horizontal axis gives the measurement time since 23 of January till 15 of April of 2014.

Fluctuations of the cosmic background are determined by fluctuations in atmospheric pressure, which can be seen in this figure in the form of anticorrelation (an increase in atmospheric pressure increases the density of atmospheric air, which shields better atmospheric muons). First of all, fluctuations in time cannot give the effect of periodic variations in space and simulate an oscillation effect. The average deviation in atmospheric pressure fluctuations is ± 2%. In our experiment, the expansion of the ON-OFF count distribution normalized to statistical error (Figure 34 (bottom), page 19) is (7 ± 4)% and is determined by the fluctuations in the cosmic space and the change in reactor power from cycle to cycle within 2-3%. Thus, this affects the value of the statistical error of the experiment, i.e. increases it by (7 ± 4)%, but cannot in any way form the effect of periodic oscillations with an amplitude of 0.26 ± 0.08. It is 7% of 0.08..

An important role is played by the frequency of switching on and off the reactor and the movement of the detector. The working regime of the reactor and the detector movement scheme were shown in the commented article in Figure 34 (top), page 19: «The scheme of reactor operation and detector movements is shown in Fig. 34 at the top. A reactor cycle is 8-10 days long. Reactor shutdowns are 2-5 days long and usually alternates (2-5-2-...). The reactor shutdowns in summer for a long period for scheduled preventive maintenance. The movement of the detector to the next measuring position takes place in the middle of reactor operational cycle. Then the measurements are carried out at the same position until the middle of the next cycle.»

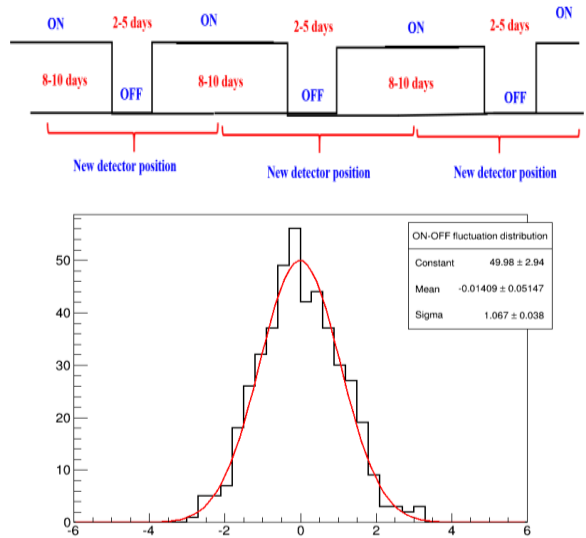

FIG. 34. Top - scheme of detector operation and detector movements; bottom - the distribution of deviations from average value of correlated events rates differences (ON-OFF) normalized on their statistical uncertainties.

Clearly, a high frequency of reactor power switching and short interval of working reactor is most preferred. In the Neutrino-4 experiment, there is both a high frequency and a short interval compared to the PROSPECT and STEREO experiments, so the problems discussed relate rather to these experiments. Finally, we cannot ignore the fact that in our experiment there were 87 reactor turns on and off, while the total time in the ON state was 720 days, in the OFF state 417 days.

The sentence astonishes: "As discussed on page 8 in Ref. [10], in an earlier analysis, a fit performed using only 278 days of reactor-off data was found to yield an oscillation parameter at ∼2.6$\sigma$ C.L."
Apparently, there is a typo 2.6$\sigma$ instead of 2.8$\sigma$. The essence of the question is not clear. The fact is that for a complete series of measurements, the statistics were increased by 1.5 times. In this regard, the accuracy was to reach 3.4sigma. The article specifies the range 3.2-3.5sigma. The level of reliability of observation of the effect of oscillations increases in proportion to the root of the measurement time. The essence of the question is not clear.

Finally, the most important background problem is the correlated background due to fast neutrons from the reactor, which can simulate IBD reaction (Figure 13, page 10).

*The problem of fast neutrons*

*False event*

*Neutrino event*

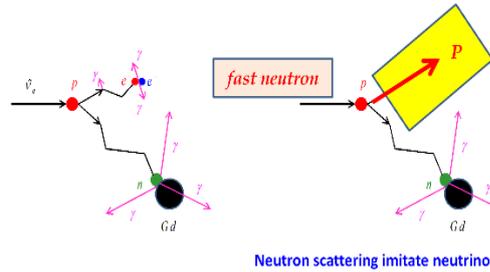

FIG. 13. The Illustration of a background problem from fast neutrons.

In this regard, before installing the detector into passive shielding, it was carefully measured the change in the background of fast neutrons when the reactor was turned on. It was shown that even at a distance of 5 m (biological reactor shielding), the background of fast neutrons hardly increases by 3% (Figure 5, page 6).

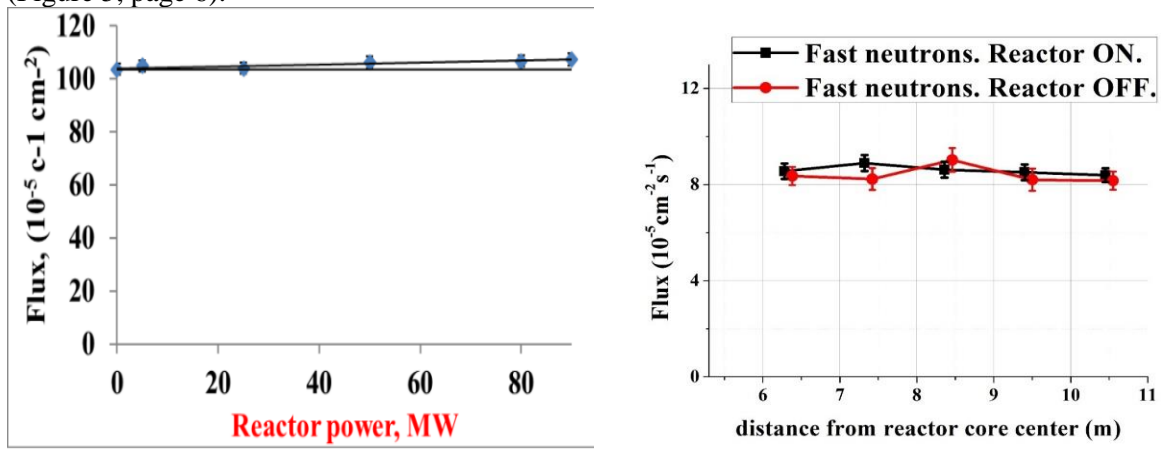

FIG.5. Left – plot of neutron flux (near the reactor wall, i.e. at distance 5.1 m from the reactor core) as a function of reactor power. Right - Fast neutron background at various distances from the reactor core measured with the detector of fast neutrons inside passive shielding. The detector of fast neutrons was placed on top of the neutrino detector and was moved with it.

The fast neutrons background in passive shielding turned out to be an order of magnitude less, so the estimate of the correlated background from the reactor does not exceed 0.3%. The background of fast neutrons is determined by neutrons that arise in the shielding of the detector and reactor from the interaction of muons with matter.

Thus, the ON-OFF count difference is determined by the neutrino flux from the reactor.

Finally, it was shown experimentally that the background of fast neutrons cannot simulate the effect of oscillations. In the article, which is commented, there is section 20 about possible systematics effects:

"1. Study of possible systematic effects was performed using a background of fast neutrons created by cosmic rays. In order to study systematic effects one have to turn off antineutrino flux (turn off the reactor) and perform the same analysis of collected data. That procedure has sufficient precision since even spectrum of recoil protons has shape very close to positron spectrum in antineutrino registration (fig. 38)

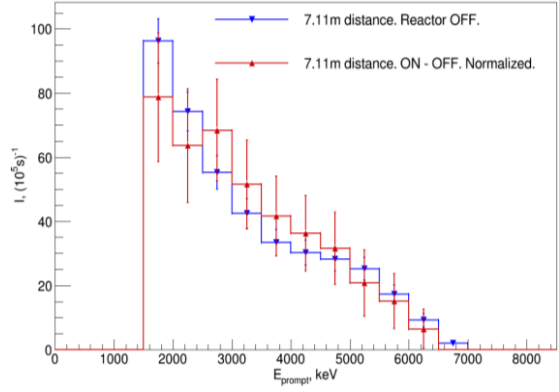

FIG. 38 Spectra of recoil protons from neutron scattering and positrons in antineutrino registration.

The result of that analysis is shown in fig.39 and it indicates the absence of oscillations in researched area. Correlated background (fast neutrons from cosmic rays) slightly decreases at farther distances from reactor due to inequality of concrete elements of the building, which comes out as linear decrease (red line) in fig. 39 (top). The deviation of results from linear law that is showed in fig.39 (bottom) cannot be used to explain the observed oscillation effect. Therefore, we can conclude that the apparatus does not produce systematical errors.

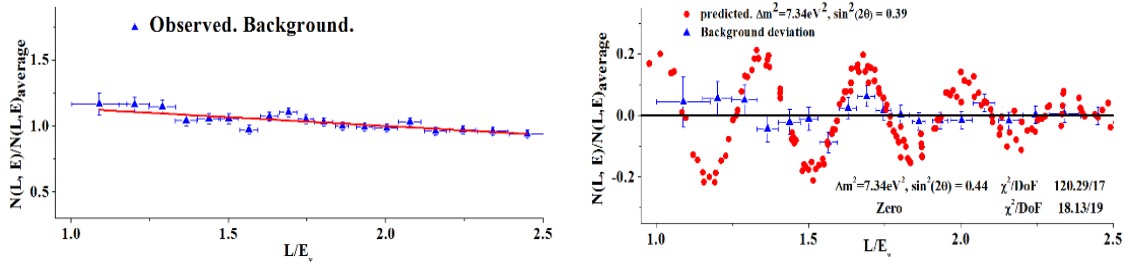

FIG.39. Analysis of data obtained with turned off reactor carried out to test on possible systematic effects: left-data analysis using coherent summation method; right - dots corresponds to deviation of expected effect from the unit, triangles - deviation of background from the linearly decreasing trend which is significantly smaller."

In response to section 4 of your comment on comparing the two methods (L/E or L,E separately approaches), we would like to conclude that we consistently use both methods as indicated in section 19 on page 20: «As previously noted, the effect of oscillations can be revealed from the construction of the dependence of the experimental ratio $N_{ik}L_k^2/K^{-1}\sum N_{ik}L_k^2$ as function from L/E. Coherent sum of data with same L/E allows to demonstrate oscillation effect directly. Method $\Delta\chi^2$, using earlier for comparison E,L matrix with calculated one, allows to find the presence of oscillations and identifies optimal parameters. Using these optimal parameters, we construct an experimental ratio $N_{ik}L_k^2/K^{-1}\sum N_{ik}L_k^2$ as dependence from L/E and compare it with calculated dependence. Method $\Delta\chi^2$ is used again and optimality of parameters is checked.»

**Conclusion.**

It seems that the questions that arise in the POSPECT and STEREO experiments are addressed to the Neutrino-4 experiment, although solving these problems were provided for by the setting of our experiment, have been implemented and tested. The advantage of the Neutrino-4 experiment is:
1. Mobile detector.
2. A method of relative measurements that does not require measurement of the absolute efficiency of the detector.
3. Duplicate measurements at the same distances by 8 different rows of detector.
4. Using spectrum independent ratio $R_{ik}^{\exp} = N_{ik}L_k^2/K^{-1}\sum N_{ik}L_k^2$ (but not for $ON - OFF = N(E_i, L_k)$) to avoid problem with spectrum of reactor antineutrino.
5. A large range of distances (6-12m) that allows successful data analysis based on L/E ratio. Which is crucial for directly demonstrating the effect of oscillations.

**Summing up, we would like to note that the effect of oscillations is manifested using three processing methods.**

    **1. $\Delta\chi^2$ method at plane ($\sin^2 2\theta_{14}, \Delta m_{14}^2$),**

    **2. Coherent summation method by variable L/E,**

    **3. Analysis of distribution $R_{ik}^{\text{exp}}$ as opposed to normal distribution due to the effect of oscillations**

**Perspectives and challenges of Neutrino-4, STEREO and PROSPECT experiments.**

We would like to say that the solution to the problem of the existence of a sterile neutrino cannot be made by detailed disassembles of the Neutrino-4 experiment at the level of statistical reliability of the $3\sigma$.

Real action would be to increase the accuracy of the PROSPECT experiments, in which to date statistics have been collected in only 33 days with the reactor operating at 85MW and 28 with the reactor powered off, and STEREO with 179 days with the reactor operating at 58.3MW and 235 days with the reactor turned off. It is quite obvious that with the continuation of measurements, the sensitivity of these experiments can be doubled and reach the sensitivity compared with the Neutrino-4 experiment, where the measurement time was 720 days with the reactor operating (90MW) and 417 with stopped.

| Experiment | Days with reactor ON | Days with reactor OFF | S/B ratio | Number of events, d$^{-1}$ |
|---|---|---|---|---|
| Neutrino-4 | **720** (90 MW) | **417** | 0.5 | 223 (6-9 m) |
| PROSPECT | 33 (85 MW) | 28 | 1.3 | 771 (7-9 m) |
| STEREO | 179 (58 MW) | 235 | 1.1 | 366 (9 – 11m) |

In the Neutrino-4 experiment, the continuation of the data taking on the existing experimental setup makes it difficult to increase the accuracy of the experiment significantly. Therefore, the experiment plans to create a new laboratory and a new detector with a sensitivity of 3 times higher in order to achieve a confidence level of $5\sigma$. We assume in the near future sensitivity of PROSPECT and STEREO experiments will be improved due to continuation of data taking. We hope that our result will be confirmed on these experiments.

**Performing this promising task will answer the question about the existence of sterile neutrinos at a new level of accuracy. Confirmations of sterile oscillations by the three experiments is significantly important.**